\documentclass[prl,twocolumn,superscriptaddress]{revtex4} 

\usepackage[utf8]{inputenc} 
\usepackage{graphicx}

\usepackage{color}

\begin{document}

\title{Spin waves in zigzag graphene nanoribbons and the stability of edge ferromagnetism}

\author{F. Culchac} 
\affiliation{Instituto de F\'isica, Universidade Federal Fluminense, 24210-346 Niter\'oi, RJ, Brazil} 

\author{A. Latg\'e} 
\affiliation{Instituto de F\'isica, Universidade Federal Fluminense, 24210-346 Niter\'oi, RJ, Brazil}

\author{A. T. Costa}\email{antc@if.uff.br} 
\affiliation{Instituto de F\'isica, Universidade Federal Fluminense, 24210-346 Niter\'oi, RJ, Brazil}
\affiliation{Department of Physics and Astronomy, University of California, Irvine, California 92697, USA}

\begin{abstract}

We study the low energy spin excitations of zigzag graphene 
nanoribbons of varying width. We find their energy dispersion 
at small wave vector to be dominated by antiferromagnetic 
correlations between the ribbon's edges, in accrodance with 
previous calculations. We point out that spin wave lifetimes
are very long due to the semi-conducting nature of the electrically 
neutral nanoribbons. However, application of very modest gate voltages 
cause a discontinuous transition to a regime of finite spin wave lifetime.
By further increasing doping the ferromagnetic alignments along the edge 
become unstable against transverse spin fluctuations. This makes
the experimental detection of ferromagnetism is this class of 
systems very delicate, and poses a difficult challenge to the possible uses 
of these nanoribbons as basis for spintronic devices.

\end{abstract} 

\maketitle

Graphene is being hailed as the big promise for nanoelectronics and 
spintronics. Its unique transport properties are expected to play a 
fundamental role in the develpment of new 
technologies~\cite{wakabayashi:1996,okada:2001,applications:palacios:2009}. 
New physics is also emerging from the interplay between low dimensionality, 
a bipartite lattice and electron-electron interaction. One of the most striking 
properties of graphene nanoribbons is the possibility of spontaneous 
magnetization~\cite{louie:nature:2006,louie:prl:2006,harrison:graphene:2007}.
This, combined with the long spin-coherence times of electrons propagating 
across graphene, indicates that this system is a strong candidate for future 
spintronics applications. 

The ground state properties of magnetic graphene nanoribbons have 
been extensively explored by a variety of methods.
Recent works have investigated the properties of 
\textbf{static} excited states based on adiabatic 
approximations~\cite{katsnelson:prl:2008,moon:prb:2009}.
This approach has been employed to 
describe the lowest-lying excitations of magnetic metals with relative 
success. However, it is well known that it misses important features of the 
excited states, such as its finite lifetime.
This arises due to the coupling 
between spin waves and Stoner excitations, a distinctive feature of itinerant 
magnets. Moreover, these recent investigations of excited states seem to have 
disregarded the antiferromagnetic coupling between the magnetizations on 
opposite edges of graphene nanoribbons. As we shall see, this leads to an 
incorrect prediction concerning the wave vector dependence of low energy spin 
excitations. This has already been demonstrated more than a decade ago in the 
seminal work by Wakabayashi \textit{et al.}~\cite{wakabayashi:1998}. 
Those authors used an itinerant model to describe the $\pi$ electrons in 
graphene nanoribbons of various widths. They showed clearly the presence
of a linear term in the spin wave dispersion relation for small wave vector. 
  
One interesting feature of magnetic graphene nanoribbons is that the spins along each 
border are ferromagnetically  coupled to each other, but there is an 
antiferromagnetic exchange coupling between the two opposite borders.  This 
coupling is mediated by the conduction electrons, and decreases as the ribbon 
width is increased.  Thus, it may appear, at first sight, that this 
antiferromagnetic coupling should be unimportant in wide ribbons. It has been 
shown, however, that this coupling is extremely long ranged in graphene and 
other related materials~\cite{meu:mmgraphene:2009,meu:cnt:2005,meu:njp:2008,meu:cnt:2008,meu:carbon:2009}.
Thus, even in rather wide nanoribbons this coupling asserts itself, as we shall see.   

We describe the electrons in graphene using a Hubbard model,
\begin{equation} 
H=\sum_{ij}\sum_\sigma t_{ij}c^\dagger_{i\sigma}c_{j\sigma} + U\sum_i n_{i\uparrow}n_{i\downarrow} 
\end{equation}
where $t_{ij}$ are hopping integrals ($i\neq j$) and on-site energies ($i=j$), 
$U$ is the effective intra-atomic Coulomb interaction and $c^\dagger_{i\sigma}$ 
creates one electron at the atomic state at site $i$ with spin 
$\sigma$. Here we only consider nearest neighbor hoppings. We took $U=2$ eV ($\approx 0.77t$), as in Ref.~\cite{macdonald:graphene:2009:a}. 
This model provides a good qualitative description of $\pi$ electrons in graphene, as well as the magnetic effects deriving from the screened Coulomb interaction.
The magnetic ground state is described self-consistently within a 
mean-field approximation. We impose local charge neutrality on every atom in the ribbon and determine the magnetic moment of each atom in the unit cell individually. We find that the magnetic moment of the edge atoms is $0.24\mu_B$ and decays rapidly towards the center of the ribbon, in close agreement with calculations based on density functional theory~\cite{katsnelson:prl:2008}. Notice that we consider the effective Coulomb interaction to be active in every atom in the system. The fact that the magnetization is essentially localized at the edges emerges naturally from our self-consistent treatment.

The spin 
excitations are extracted from the properties of the transverse dynamic susceptibility, 
\begin{equation}
 \chi^{+-}_{ij}(t)=-i\theta(t)\left\langle\left[S^+_i(t),S^-_j(0)\right]\right\rangle, 
 \end{equation} 
where $S^+=a^\dagger_\uparrow a_\downarrow$ and $S^-=(S^+)^\dagger$ are the spin raising and lowering operators.  
By treating the Coulomb interaction term within a random phase approximation we obtain a closed equation of 
motion for $\chi^{+-}(\Omega)$ (the Fourier transform of $\chi^{+-}(t)$) in terms of the mean-field susceptibility $\chi^{(0)+-}(\Omega)$~\cite{copinho},
\begin{equation}
\chi^{+-}(\Omega) = \left[ I + U\chi^{(0)+-}(\Omega)\right]^{-1}\chi^{(0)+-}(\Omega), 
\label{dyson}
\end{equation}
where $\chi^{+-}$ and $\chi^{(0)+-}$ are matrices comprising all magnetic sites in the system and $I$ is the identity matrix with the same dimension as $\chi^{+-}$ and $\chi^{(0)+-}$. The notation used in the equation above is schematic. A more precise statement on the form of the calculated susceptibility is given below.

The spectral density  
 \begin{equation} 
 A_i(\Omega) = -\Im\chi_{ii}(\Omega), 
 \label{spectral_density}
 \end{equation} 
where $\Im$ denotes the imaginary part, may be interpreted as the density of states of magnons in the system. The dynamic susceptibility just described is the response of the system to an externally applied field of frequency $\Omega$ transverse to the ground state magnetization direction; spin waves appear as peaks in the spectral density.

\begin{figure}
	\centerline{\includegraphics[clip,width=0.8\columnwidth]{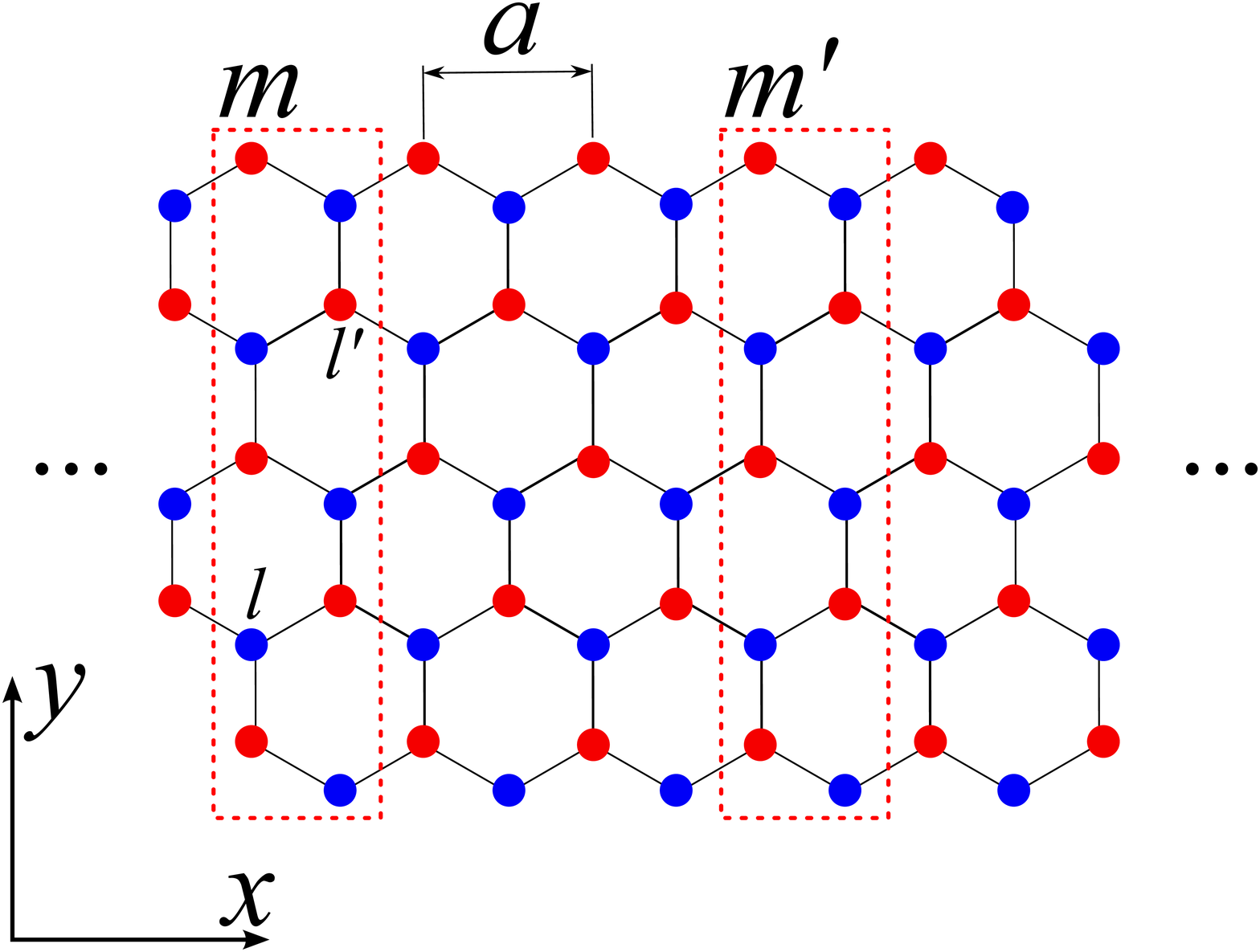}}
	\caption{(Color online) Schematic depiction of the zigzag nanoribbon's geometry. The dotted lines encircle two arbitrary unit cells, labeled $m$ and $m'$. The indices $l$ and $l'$ refer to atoms inside each unit cell (as in equation~\ref{bloch_wannier}).}
	\label{unit_cell}
\end{figure}

The graphene nanoribbons we study have translation symmetry along the ribbon length (denoted here by $x$), but not along the  
ribbon width ($y$). It is convenient to define a mixed Bloch-Wannier basis to describe the electronic states, 
 \begin{equation} 
 c_l(k)=\frac{1}{\sqrt{N}}\sum_{m}e^{imka}c(x_{m},y_l),
 \label{bloch_wannier} 
 \end{equation} 
 where $c(x_m,y_l)$ is the annihilation operator for a Wannier state at a site $l$
 in unit cell $m$, $a=\sqrt{3}a_0$ is the distance between unit cells along the ribbon 
 length and $a_0\approx 1.42$~\AA\ is the carbon-carbon distance. In this representation, the transverse dynamical susceptibility
$\chi^{+-}_{ll'}(\Omega;k)$
is a matrix, where $l,l'$ label sites within a unit cell; each element of such matrix is a function of the wave vector $k$ along the length 
of the ribbon, as well as of the energy $\Omega$. The unit cell is depicted in Fig.~\ref{unit_cell}

% \section{Results}

\begin{figure}
	\centerline{\includegraphics[clip,width=\columnwidth]{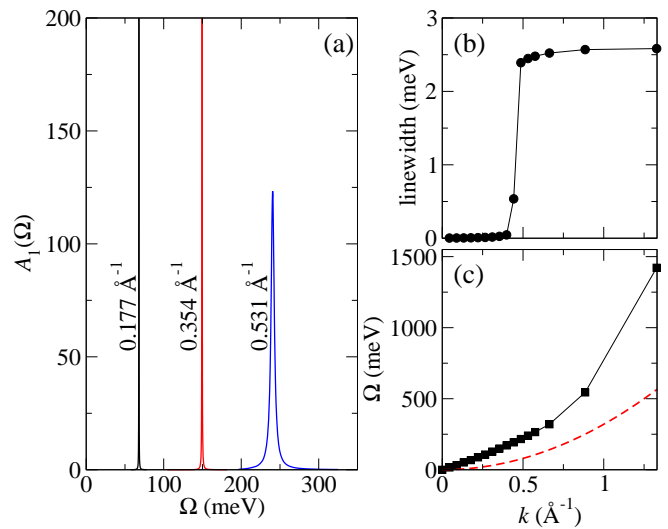}}
	\caption{(Color online) (a) Spectral density associated with spin waves, projected on the 
	``up'' edge, for a neutral ribbon 8 atoms wide, for selected wave vectors 
	(indicated in the figure); (b) linewidth as a function of wave vector; 
	(c) spin wave dispersion relation deduced from the peaks of the spectral 
	density (squares). The dashed curve is a plot of the quadratic dispersion 
	relation found in adiabatic calculations~\cite{katsnelson:prl:2008}.}
	\label{fig1}
\end{figure}

We start by discussing the spectral density $A_i(k;\Omega)$ 
for a nanoribbon of fixed width. In Fig.~\ref{fig1} we plot the spectral 
density as a function of $\Omega$ for fixed values of wave vector $k$, for a 
ribbon with 16 atoms in the cross-section. The main contribution for the 
excitations should come from the edges, where most of the magnetic moment of 
the system is concentrated. Thus, we only need to plot the spectral density 
projected on the ``up'' edge, that we label as $i=1$ (the spectral density at 
the ``down'' edge has similar behavior). Spin 
wave energies increase with wave  vector, as usual, but the $k$ dependence is 
not quadratic, as would be expected from a simple ferromagnet. 

A plot of the dispersion relation deduced from the peak positions (Fig.~\ref{fig1}(c)) shows that  the dispersion is linear quite far out into the 
Brillouin zone (20\% of the zone boundary), and in fact is quasi linear at 
large wave vectors as well. This may be understood if we map the spin degrees 
of freedom of this system onto a simple effective spin model, as illustrated in 
ref.~\cite{wakabayashi:1998}. 

Calculations of spin excitation energies based on an adiabatic approach have 
been reported recently~\cite{katsnelson:prl:2008}. They find a quadratic 
energy-wave vector dispersion relation with a spin wave exchange stiffness of 
320 meV\AA$^2$. We plot this dispersion relation in Fig.~\ref{fig1}(c) for comparison. Although the energies found using the 
adiabatic approach are of the same order as those obtained via dynamical 
calculations, the discrepancy between the dependencies on wave vector is 
remarkable.

In Fig.~\ref{fig1}(a) we see that the spin wave peaks are extremely 
narrow for small wave vectors, indicating a very large spin wave lifetime.
This is compatible with the existence of a threshold for Stoner excitations of 
the order of the semiconducting gap in these ribbons. Only spin waves with 
energies equal to or larger than this gap are damped.
This means that the low energy spin dynamics (represented by long wavelength 
spin waves) may be well 
described by effective localized spins hamiltonians, but as the wavelength of
the excitation become smaller the itinerant nature of the system reveals itself. 
Thus, a simple ferromagnetic Heisenberg hamiltonian is clearly \textbf{not}
the appropriate model to describe the spin degrees of freedom of this fascinating
system.

%\section{Doping}

\begin{figure}
	\centerline{\includegraphics[width=\columnwidth]{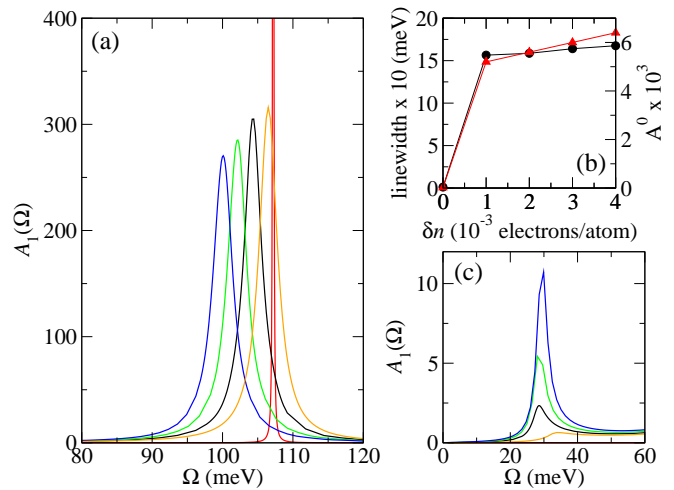}}
	\caption{(Color online) (a) Spectral densities for $q=0.266$~\AA$^{-1}$ and different 
	doping levels (0, 1, 2, 3 and 4 $\times 10^{-3}$ electrons/atom) for a ribbon 8 atoms wide. 
	Larger doping means smaller energy and smaller peak height. 
	(b) Linewidth as a function of doping (solid circles, scale on the left) 
	and density of Stoner modes $A^0$ at the spin wave energies (solid triangles, 
	scale on the right). (c) Spectral 
	density at low energies for the same wave vector and doping levels as 
	panel (a). Notice that the spectral density is absolutely flat in this 
	region for zero doping.}
	\label{chi_x_doping}
\end{figure}

One very attractive feature of graphene is the possibility of controlling its 
carrier density by electrostatic gating. In the present case this feature 
opens up a very exciting possibility: by controlling the electron density we 
may be able to tune the relaxation time of spin excitations in graphene. As 
shown above, long wavelength spin waves are essentially undamped in 
electrically neutral graphene ribbons. In fig.~\ref{chi_x_doping} we show that very 
modest changes in the electron density can induce rather large damping, 
reducing considerably the relaxation time for spin excitations and shifting 
their energies. The origin 
of this damping is simple to grasp: the density of Stoner modes is very small 
at small energies in undoped graphene ribbons due to the fact that the density 
of states $\rho$ near the Fermi level $E_F$ is zero (the antiferromagnetic, undoped 
nanoribbon is semiconducting). As the density of states is increased by 
the gate voltage, $\rho$ is increased for energies close to $E_F$, giving rise 
to a significant enhancement of the density of Stoner modes. As it is well 
known~\cite{copinho,bechara:FeW:2002,tang:plihal:mills,meu:FeW:prb:2003}, 
spin wave damping in itinerant systems occurs through 
the decaying of magnons into Stoner modes, a mechanism very similar to the 
Landau damping of plasmons in metals. In Fig.~\ref{chi_x_doping}(b) we show 
how the density of Stoner modes at the spin wave energies (given by the spectral 
function $A^0$ associated with the non-interacting susceptibility $\chi^{(0)+-}$) 
is enhanced by increasing electron density.

It is also clear from Fig.~\ref{chi_x_doping} that the extra damping is accompanied 
by a shift in the spin wave energy. Once again, this is related to the 
enhancement of the density of Stoner modes, via the Kramers-Kr\"onig relation. 
The non-interacting susceptibility $\chi^{(0)+-}$ enters in the denominator of the 
dynamic susceptibility $\chi^{+-}$, as indicated in Eq.~\ref{dyson}. Its imaginary 
part is responsible for the finite lifetime of spin waves in itinerant 
magnets; its real part produces a shift in the spin wave frequencies, in much 
the same way as dissipative forces shift the natural frequency of mechanical 
oscillators. Thus, enhancement of damping also implies a larger frequency shift.

There is another facet to the onset of strong spin wave damping in graphene 
nanoribbons. Spin excitations with infinite (or extremely long) lifetimes are 
associated with strongly localized spins, whereas strongly damped spin waves 
are found in systems where magnetism is itinerant in nature. It is very rare 
that one system can be tuned to be either a localized or an itinerant magnet 
with the change of a single parameter, easily accessible experimentally. It is 
an extremely exciting prospect that this kind of control is available in 
graphene nanoribbons.

% Instability

The lifetime of spin waves in zigzag graphene nanoribbons can be dramatically reduced,
as we just saw, by very modest doping (as small as $10^{-3}$ electons/atom). By further 
increasing doping we notice that the ferromagnetic alignment along the borders becomes 
unstable. A sign of this instability is the appearance of a very soft spin wave mode as
doping increases, as can be seen in Fig.~\ref{chi_x_doping}(c).

\begin{figure}
	\centerline{\includegraphics[width=\columnwidth]{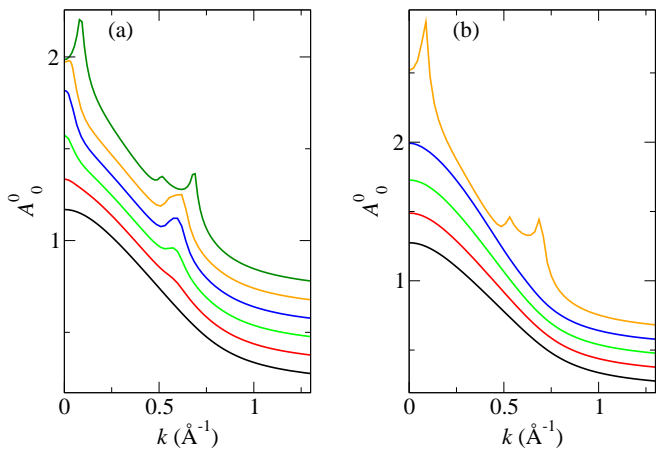}}
	\caption{(Color online) The zero frequency mean-field spectral density $A^0_0(k)$ for 
	different doping levels and two values of the effective Coulomb interaction, 
	$U=2$~eV (a) and $U=1.5$~eV (b). The curves have been displaced vertically 
	for the sake of clarity. In panel (a) the doping levels are 0, 2, 3, 4, 5 and 
	10 m$e$/atom; in panel (b) the doping levels are 0, 5, 10, 15 and 20~m$e$/atom. 
	Doping increases from bottom to top. }
	\label{chi0_x_k}
\end{figure}

The instability of the ferromagnetic alignment can be confirmed by
the behavior of the mean-field transverse susceptibility at zero frequency, 
as a function of wave vector, 
$A^0_0(k)\equiv \chi^{(0)+-}(k,\Omega=0)$~\cite{edwards:1979}. In a stable ferromagnetic 
system $A^0_0(k)$ has a single maximum at $k=0$, as 
illustrated by the zero doping curve in figure~\ref{chi0_x_k}(a).
As doping increases, a peak develops close to $k=0$, until, at large
enough doping (in this case, 0.01 electrons/atom), a pronounced
maximum appears at a finite value of $k$. The existence of peaks in  
$A^0_0(k)$ at finite values of $k$
means that the true ground state of this system is not ferromagnetic along the 
edges, but most probably a spin density wave characterized by those finite 
wave vectors.

One virtue of our simple model is that we can tune parameters and explore
various behaviors. By changing the strength $U$ of the Coulomb interaction we
noticed that the doping level at which the
instability appears changes. This is illustrated in 
Fig.~\ref{chi0_x_k} for two different values of $U$; for $U=2$~eV
the ferromagnetic alignment becomes unstable at $\sim 5\times 10^{-3}$~electrons/atom,
whereas for $U=1.5$~eV it is stable for doping levels as large as 
$1.5\times 10^{-2}$~electrons/atom.
It would be interesting to build a $U\times\delta n$ phase diagram, 
but our intention here is to point out the dependence and the general trend.

The stability analysis we performed is complementary to that presented
in references \cite{macdonald:graphene:2009:a,macdonald:graphene:2009:b,sawada:2009},
where energy differences between collinear and non-collinear configurations 
in the direction transverse to the ribbon were analized. To the best of
our knowledge, ours is the first analysis that take into account the 
possibility of non-collinear ordering \textbf{along} the ribbon's edges.

We have investigated wider nanoribbons (up to 32 atoms wide, although this is 
only limited by computer time). The most important effects of increasing the 
ribbon width are i) the enhancement of a quadratic contribution to the spin wave 
dispersion relation (due to the partial suppression of antiferromagnetic 
coupling between edges) and ii) the reduction of the doping range (for fixed $U$)
within which ferromagnetism along the borders is stable.  

In conclusion, we have shown that the lifetimes of spin excitations in
zigzag graphene nanoribbons can be tuned by the application of modest gate
voltages. This allows, at least in principle, electrical control of the magnetic
relaxation rate. We have also demonstrated that there is a sharp transition between 
the character of the spin excitations in neutral and doped nanoribbons: while
in neutral ribbons the long wavelength excitations have essentially infinite lifetime 
(a feature shared with magnetic insulators), any amount of doping, however 
small, leads to finite lifetimes (as in magnetic metals). Finally,
we showed that further increasing doping makes the ferromagnetic alignment 
unstable against transverse spin fluctuations, a fact that should be carefully
invesigated if these systems are to be used in technological applications.
We are confident that our results open very exciting possibilities both in 
spintronics technology and for fundamental understanding of magnetism at the 
nanoscale. 

\begin{acknowledgments}
	We are extremely grateful to R. B. Muniz, D. L. Mills and M. S. Ferreira for discussions and critical readings of the manuscript. The research of A.T.C. and A.L.  was supported by CNPq and FAPERJ (Brazil). F.C. acknowledges a scholarship from CNPq (Brazil). We also acknowledge partial financial support from the National Institute of Science and Technology - Carbon Nanomaterials (INCT-NT).
\end{acknowledgments}

% \bibliography{graphene_bib}

\end{document}